# Comparing journals from different fields of Science and Social Science through a JCR Subject Categories Normalized Impact Factor

P. Dorta-González [a*], M.I. Dorta-González [b]

[a] Departamento de Métodos Cuantitativos en Economía y Gestión, Universidad de Las Palmas de Gran Canaria, Gran Canaria, España; [b] Departamento de Estadística, Investigación Operativa y Computación, Universidad de La Laguna, Tenerife, España.

**ABSTRACT**

The journal *Impact Factor* (IF) is not comparable among fields of Science and Social Science because of systematic differences in publication and citation behaviour across disciplines. In this work, a decomposing of the field aggregate impact factor into five normally distributed variables is presented. Considering these factors, a Principal Component Analysis is employed to find the sources of the variance in the *JCR subject categories* of Science and Social Science. Although publication and citation behaviour differs largely across disciplines, principal components explain more than 78% of the total variance and the average number of references per paper is not the primary factor explaining the variance in impact factors across categories. The Categories Normalized Impact Factor (CNIF) based on the JCR subject category list is proposed and compared with the IF. This normalization is achieved by considering all the indexing categories of each journal. An empirical application, with one hundred journals in two or more subject categories of economics and business, shows that the *gap* between rankings is reduced around 32% in the journals analyzed. This *gap* is obtained as the maximum distance among the ranking percentiles from all categories where each journal is included.

# 1. Introduction

The *Impact Factor* (IF) published in the *Journal Citation Reports* (JCR) by Thomson Reuters is defined as the average number of citations to each journal in a current year to 'citable items' published in that journal during the two preceding years. Since its formulation (Garfield, 1972), the IF has been criticized for some arbitrary decisions involved in its construction. The definition of 'citable items' (articles, notes, and reviews), the focus on two preceding years as representation of impact at the research front, etc., have been discussed in the literature (Bensman, 2007) and many possible modifications and improvements have been suggested (Althouse et al., 2009). In response, Thomson Reuters has added the *Five-year Impact Factor*, the *Eigenfactor Score*, and the *Article Influence Score* (Bergstrom, 2007) to the journals in the online version of the JCR, in 2007 (see Bornmann & Daniel, 2008, for a review in citation measures). While the extension of the IF to a five-year time window is direct, the other recent measures can be considered too complex (Waltman & Van Eck, 2010) and they do not solve the problem of comparing journals from different fields of science.

The problem of field-specific differences in citation impact indicators comes from institutional research evaluation (Leydesdorff & Opthof, 2010a; Opthof & Leydesdorff, 2010; Van Raan et al., 2010). Citation distribution varies with fields of science and, in some cases, across specialties within fields (Dorta-González & Dorta-González, 2010, 2011a,b). However, institutes are populated by scholars with different disciplinary backgrounds and research institutes often have among their missions the objective of integrating interdisciplinary bodies of knowledge (Leydesdorff & Rafols, 2011; Wagner et al., 2011).

There are statistical patterns which are field-specific and allow the normalization of the IF. Garfield (1979a,b), proposes the term 'citation potential' for systematic differences among fields of science based on the average number of references per paper. For example, in the biomedical fields long reference lists with more than fifty items are common, but in mathematics short lists with fewer than twenty references are the standard. These differences are a consequence of the citation cultures, and can be expected to lead to significant differences in the IF across fields of science because the probability of being cited is affected. The fractionally counted IF corrects these differences in terms of the sources of the citations (Leydesdorff & Bornmann, 2011; Moed, 2010; Zitt & Small, 2008). Using fractional counting, a citation in a citing paper containing $n$ references counts $1/n$, instead of 1, as is the case with integer counting.

In relation to the source-normalization, Zitt & Small (2008) propose the *Audience Factor* (AF) using the mean of the fractionally counted citations to a journal. This mean is then divided by the mean of all journals included in the Science Citation Index. Similarly, Moed (2010) divides a modified IF (with a window of three years and a different definition of citable items) by the median number of references in the Scopus database. He proposes the resulting ratio as the *Source Normalized Impact per Paper* (SNIP) which is now in use as an alternative to the IF in the Scopus database (Leydesdorff & Opthof, 2010b). The *Scimago Journal Ranking* (SJR) considers the prestige of the citing journals (González-Pereira et al., 2011), and even though this is useful for the ranking of journals, the value of the indicator is difficult to interpret (Waltman et al., 2011).

Another important source of variance between fields is related to the dissemination channel of the research activity results. For example, researchers in social sciences and humanities publish more in books than in journals, and researchers in computer science publish their results more in conference proceedings than in journal articles (Chen & Konstan, 2010; Freyne et al., 2010). Differences between fields citations are caused mainly by the different ratio of references to journals included in the JCR as opposed to references to 'non-source items' (e.g., books), whereas differences in the lengths of reference lists are mainly responsible for inflation in the IF (Althouse et al., 2009).

Most efforts to classify journals in terms of fields of science have focused on correlations between citation patterns in core groups assumed to represent scientific specialties (Leydesdorff, 2006; Rosvall & Bergstrom, 2008 and 2010). Indexes such as the *JCR subject category list* accommodate a multitude of perspectives by listing journals under different groups (Pudovkin & Garfield, 2002; Rafols & Leydesdorff, 2009). In this sense, Egghe & Rousseau (2002) define the *Relative Impact Factor* (RIF) in a similar way as the IF, taking all journals in a category as a meta-journal. This indicator in the JCR is called Aggregate Impact Factor (AIF). Furthermore, normalizations based on the journal ranking in the category (Sombatsompop & Markpin, 2005), and the maximum value of the IF jointly with the median in the category (Ramírez et al., 2000) have been proposed.

However the positions of individual journals on the merging specialties remain difficult to determine with precision and some journals are assigned to more than one category. This is the case of Science, Nature, and the Proceedings of the National Academy of Science of the USA (PNAS), which are examples of high prestige multidisciplinary journals. Many others journals cover two or more specialties. Therefore, journals cannot easily be compared, and classification systems based on citation patterns hence tend to fail. When a journal publishes work from several subject categories, its performance may be better when seen from the standpoint of one subject category than from the other (multidisciplinary effect).

The categories in the JCR were created in order to compare journals within the same category. However, what is to be done when a journal is included in more than one category? In this work, a normalization process considering all journals in the indexing categories is proposed. In order to compare the normalized impact indicator with the IF, an empirical application, with one hundred journals in two or more subject categories of economics and business, is presented.

In addition to the average number of references and the ratio of references to journals included in the JCR, there exist some other sources of variance between fields. In this work also we decompose the aggregated impact factor into five main sources of variance and calculate them in all the categories of the JCR. Out of the five main sources there are three new factors: the field growth, the ratio of JCR references to the target window, and the proportion cited to citing items.

**2. Decomposing the Aggregate Impact Factor of a field into its main components**

*2.1 Impact Factor of a journal*

A journal impact indicator is a measure of the number of times that articles published in a census period cite articles published during an earlier target window. The IF reported by Thomson Reuters has a one year census period and uses the two previous years as the target window.

As an average, the IF calculation is based on two elements: the numerator, which is the number of citations in the current year to any items published in a journal in the previous two years, and the denominator, which is the number of 'citable items' published in the same two years (Garfield, 1972). Journal items include 'citable items' (articles, notes, and reviews), but also letters, corrections and retractions, editorials, news, and other items.

Let $A_t^i$ be the number of citable items in journal *i* in year *t*. Let $NCited_t^i$ be the number of times in year *t* that the year *t-1* and *t-2* volumes of journal *i* are cited by journals in the JCR. Then, the Impact Factor of journal *i* in year *t* is

$$IF_t^i = \frac{NCited_t^i}{A_{t-1}^i + A_{t-2}^i}. \qquad (1)$$

*2.2 Aggregate Impact Factor (AIF) of a field*

Let *F* be the set of all journals in a specific field, where the fields are equivalent to the *JCR subject categories*. Denoting $A_t^F = \sum_{i \in F} A_t^i$ and $NCited_t^F = \sum_{i \in F} NCited_t^i$, the Aggregate Impact Factor (AIF) is the ratio between the citations in year *t* to citable items in any journal of field *F* in years *t-1* and *t-2*, and the number of citable items published in years *t-1* and *t-2*, that is,

$$AIF_t^F = \frac{\sum_{i \in F} NCited_t^i}{\sum_{i \in F} A_{t-1}^i + A_{t-2}^i} = \frac{NCited_t^F}{A_{t-1}^F + A_{t-2}^F}. \qquad (2)$$

The AIF can also be expressed as a weighted mean impact factor. Consider a formulation that assigns weights proportional to the number of citable items in the target years. The weight for journal $i$ in year $t$ is

$$f_t^i = \frac{A_{t-1}^i + A_{t-2}^i}{A_{t-1}^F + A_{t-2}^F}. \qquad (3)$$

Notice that $\sum_{i \in F} f_t^i = 1$. Then, from equations (1), (2), and (3),

$$AIF_t^F = \frac{\sum_{i \in F} NCited_t^i}{A_{t-1}^F + A_{t-2}^F} = \sum_{i \in F} \frac{NCited_t^i}{A_{t-1}^F + A_{t-2}^F} = \sum_{i \in F} \frac{A_{t-1}^i + A_{t-2}^i}{A_{t-1}^F + A_{t-2}^F} IF_t^i = \sum_{i \in F} f_t^i \cdot IF_t^i. \qquad (4)$$

*2.3 Components in the Aggregate Impact Factor of a field*

It is possible to decompose the AIF into five main variables. We will show in section 4 that these variables are normally distributed and different across fields. The variable $a_t^F$ is a measure of the field growth while the others ($r_t^F, p_t^F, w_t^F, b_t^F$) are related with the citation habits in the field.

- *Field growth rate*

A field can grow for two reasons; by incorporating new journals into the field or by publishing more items in some indexed journals. However, a field can also decrease. Let $a_t^F = A_t^F / (A_{t-1}^F + A_{t-2}^F)$ be the ratio of citable items in year $t$ with respect to those appearing in the target window. This is a measure of the field growth. Note that $a_t^F = 0.5$ when $A_t^F = A_{t-1}^F = A_{t-2}^F$. If $a_t^F > 0.5$ then the field is growing in the number of citable items. Conversely, if $a_t^F < 0.5$ then the field is reducing.

For example, if a field is growing annually around 5%, then $A_t^F = 1.05 \cdot A_{t-1}^F$, $A_{t-1}^F = 1.05 \cdot A_{t-2}^F$, and $a_t^F = (1.05^2 \cdot A_{t-2}^F) / (2.05 \cdot A_{t-2}^F) = 1.05^2 / 2.05 = 0.538$. Some others ratios are $a_t^F = 0.576$ (10%) and $a_t^F = 0.654$ (20%). On the other hand, if a field is reducing annually around 5%, then $A_t^F = 0.95 \cdot A_{t-1}^F$, $A_{t-1}^F = 0.95 \cdot A_{t-2}^F$, and $a_t^F = (0.95^2 \cdot A_{t-2}^F) / (1.95 \cdot A_{t-2}^F) = 0.95^2 / 1.95 = 0.463$.

- *Average number of references*

Let $R_t^F$ be the total number of references in journals of field $F$ in year $t$. Then $r_t^F = R_t^F / A_t^F$ is the average number of references in citable items of field $F$ in year $t$.

- *Ratio of references to JCR items*

Let $J_t^F$ be the total number of references (in items of field $F$ in year $t$) to journals in the JCR. This excludes unpublished working papers, conference proceedings, books, and journals not indexed by the JCR. Then $p_t^F = J_t^F / R_t^F$ is the ratio of number of references to number of JCR items. For example, if $p_t^F = 0.5$ then half of the references are JCR items.

- *Ratio of JCR references to the target window*

Let $NCiting_t^F$ be the total JCR references in field $F$ in year $t$ within the target window. Then, $w_t^F = NCiting_t^F / J_t^F$ is the ratio of JCR references in year $t$ within the target window. For example, if $w_t^F = 0.25$ then a quarter of the JCR references belong to the target window.

- *Proportion of cited to citing items in the target window*

If $i \in F$, most of the citations to journal $i$ came from journals within field $F$ but some of them came from journals outside field $F$. Let $b_t^F = NCited_t^F / NCiting_t^F$ be the proportion of cited to citing items in the target window. If $b_t^F > 1$ then citations received in field $F$ are more than citations produced in field $F$ (within the target window). Conversely, if $b_t^F < 1$ then citations received are less than citations produced. Therefore, the index $b_t^F$ is a measure of the field's citations exchange. For example, if $b_t^F = 1.1$ then field $F$ receives in the target window around 10% more citations than it produces.

*2.4 Decomposing the AIF of a field into components*

The Aggregate Impact Factor of field $F$ can be decomposed or factorized in the following way:

$$AIF_t^F = a_t^F \cdot r_t^F \cdot p_t^F \cdot w_t^F \cdot b_t^F. \quad (5)$$

The proof is direct, considering that $NCited_t^F$ can be expressed as:

$$NCited_t^F = A_t^F \cdot \frac{R_t^F}{A_t^F} \cdot \frac{J_t^F}{R_t^F} \cdot \frac{NCiting_t^F}{J_t^F} \cdot \frac{NCited_t^F}{NCiting_t^F} = A_t^F \cdot r_t^F \cdot p_t^F \cdot w_t^F \cdot b_t^F. \quad (6)$$

Therefore, from (2) and (6),

$$AIF_t^F = \frac{A_t^F \cdot r_t^F \cdot p_t^F \cdot w_t^F \cdot b_t^F}{A_{t-1}^F + A_{t-2}^F} = a_t^F \cdot r_t^F \cdot p_t^F \cdot w_t^F \cdot b_t^F.$$

**3. Categories Normalized Impact Factor (CNIF)**

This is a field-normalized citation impact score, where the fields are equivalent to the JCR subject categories. We compare 'actual' citation counts to 'expected' counts based on the average impact score of all JCR-indexed journals assigned to a field.

Let $F_t^1, F_t^2, ..., F_t^n$ be the subject categories where journal $i$ is indexed in year $t$. Denoting by $\bigcup F_t^j = F_t^1 \bigcup F_t^2 \cdots \bigcup F_t^n$, then

$$AIF_t^{\bigcup F_t^j} = \frac{\sum_{i \in \bigcup F_t^j} NCited_t^i}{\sum_{i \in \bigcup F_t^j} A_{t-1}^i + A_{t-2}^i}.$$

In a similar way,

$$AIF_t^{JCR} = \frac{\sum_{i \in JCR} NCited_t^i}{\sum_{i \in JCR} A_{t-1}^i + A_{t-2}^i}.$$

Let $AIF_t^{JCR} / AIF_t^{\bigcup F_t^j}$ be the normalized score of the meta-category $\bigcup F_t^j$. If $AIF_t^{JCR} = AIF_t^{\bigcup F_t^j}$ then the score is one. Scores larger than one represent aggregate impact factors in the field below the average in the JCR, while scores lower than one represent aggregate impact factors in the field above the average in the JCR.

We define the *Categories Normalized Impact Factor* of journal $i$ in year $t$ as:

$$CNIF_t^i = \frac{AIF_t^{JCR}}{AIF_t^{\bigcup F_t^j}} \cdot IF_t^i$$

Notice that if $AIF_t^{\bigcup F_t^j} > AIF_t^{JCR}$, then the score is less than one and it reduces the impact factor of that journal. Conversely, if $AIF_t^{\bigcup F_t^j} < AIF_t^{JCR}$, then the score is higher than one and it increases the impact factor of that journal.

In the particular case of a journal in only one category $F$,

$$CNIF_t^i = \frac{AIF_t^{JCR}}{AIF_t^{F}} \cdot IF_t^i$$

The CNIF has an intuitive interpretation, similar to the IF. The CNIF is a measure of the number of times that articles published in year $t$ cite, in a category-normalized proportion, articles published during the two previous years. Moreover, it is easy to calculate and allows for the comparison between fields.

**4. Empirical application**

*4.1 Materials and Methods*

The underlying bibliometric data in the empirical application was obtained from the online version of the *Journal Citation Reports* (JCR) during the first week of October 2011. The JCR database (reported by Thomson Reuters – ISI, Philadelphia, USA) is available at the website www.webofknowledge.com. The IF reported by Thomson Reuters has a one year census period and uses the two previous years as the target window.

In the JCR, journals are assigned by Thomson Reuters experts into one or more journal categories, according to cited and citing relationships with the journals in the categories (Pudovkin & Garfield, 2002). The journal categories, also referred to *subject category list,* are treated as fields and subfields of science. The 2010 Science edition contains 8073 journals classified into 174 subject categories. The 2010 Social Science edition contains 2731 journals classified into 56 subject categories. Given that these journal categories are well known, there is no reason to question the feasibility of using them in the field-normalization.

Although most journals in the JCR are included in only one edition (Science or Social Science), there are some which are included in both. This happens, for example, with nine journals included in the category 'Management', in the Social Science edition, and in the category 'Operations Research and Management Science', in the Science edition.

In the comparative analysis between the IF and the CNIF, and the estimation of the gap between rankings across categories, the five selected categories are: SS3 (BUSINESS), SS4 (BUSINESS, FINANCE), SS9 (ECON), SS29 (MANAGE), and S124 (OPER RES & MANAGE SCI). These five categories contain a total of 590 different journals. There are 490 journals in just one category, 98 journals in two categories, and 2 journals in three categories. Therefore, the number of journals in more than one category is exactly 100 (48 journals in SS3, 36 journals in SS4, 54 journals in SS9, 55 journals in SS29, and 9 journals in S124).

*4.2 Results*

- *The Aggregate Impact Factor of the JCR subject categories*

Table 1 shows the AIF for both editions of the JCR (i.e. Science and Social Science). At the bottom of Table 1 the aggregate index, the average, and the standard deviation for both editions, are also shown. Unlike the aggregate impact factor, that considers the size of the categories, in the average impact factor all the categories have the same weight. The AIF in Science is 2.920, around 58% higher than in Social Science which is 1.848.

[Table 1 about here]

There exists a great variance in the AIF within each edition. In Science, the categories with highest values are MULTIDISCIP SCI (9.707), CELL BIOL (6.453), and HEMATOL (5.310), whereas the lowest factors are for ENGN, MARINE (0.207), ENGN, PETROLEUM (0.565), and ENGN, AEROSPACE (0.628). In Social Science, the categories with the highest AIF are PSYCHIATRY (3.215), PSYCHOL, BIOL (2.682), and PSYCHOL, EXPT (2.590), whereas the lowest factors are for HIST (0.479), HIST OF SOCIAL SCI (0.623), and AREA STUDIES (0.640). Notice in Figure 1 that the AIF in Science is higher than in Social Science.

[Figure 1 about here]

- *The components of the JCR subject categories*

Table 1 shows also the components. There are important differences between categories, especially between categories from different editions.

The average growth of the JCR is 0.57 (the JCR database annual growth is around 9%). This average growth is about 0.55 (7%) in the Science edition and around 0.62 (16%) in the Social Science edition. Therefore, the Social Science edition is growing over twice as much as the Science edition. This is due to the incorporation of journals in some categories of the Social Science edition in the last few years. This has occurred in categories HOSPITAL, LEIS, SPORT & TOUR (0.97), ETHNIC STUDIES (0.81), and HIST (0.78), for example. In Science, a great variance is observed. Note the case of BIOL (0.85) in comparison to ENGN, INDUSTRIAL (0.41), for example.

The average number of references in Science is 37.18 while in Social Science it is 48.28. Therefore, a journal from a category of Social Science has on average 30% more references than a journal from a Science category. However, there exists a great variance within editions. Notice categories CELL & TISSUE ENGN (75.66), PALEONTOL (67.05), and HIST (66.28), in comparison to ENGN, MARINE (13.94), NUCLEAR SCI & TECH (19.21), and MATH (20.49), for example.

The average ratio of references to JCR items is 0.75. In Science this average is 0.80 whereas in Social Science it is 0.60. Therefore, a journal from a Social Science category has on average 20% more references to non-JCR items than a journal from a Science category. However, there exists a great variance within editions. In Science, notice categories PHYS, ATOM, MOLEC & CHEM (0.94), CELL BIOL (0.93), and CHEM, ORGANIC (0.93), in comparison to ENGN, MARINE (0.39), HIST & PHILOS OF SCI (0.44), and COMP SCI, SOFT ENGN (0.56), for example. In Social Science, note categories PSYCHOL, BIOL (0.87), PSYCHOL, EXPT (0.83), and PSYCHIATRY (0.80), in relation to HIST (0.30), and CULTURAL STUDIES (0.32), for example.

The average ratio of JCR references to the target window is 0.18. Therefore, one of each five JCR references is on average within the target window. There are few differences between editions (0.18 in Science and 0.20 in Social Science) but there exists a great variance within editions. The highest ratios are 0.45 in AREA STUDIES, 0.35 in INT RELAT, and 0.33 in ENGN, MARINE. The lowest ratios are 0.10 in PALEONTOL, and 0.11 in GEOL and MANAGE.

Finally, there exists a great variance in the proportion of cited to citing items. In Science, notice categories MULTIDISCIP SCI with 2.55, PHYS, MULTIDISCIP with 1.20, and HEMATOL with 1.18, in comparison to ENGN, MARINE with 0.28, HIST & PHILOS OF SCI with 0.30, and NURS with 0.41, for example. In Social Science, note category PSYCHOL, MATH with 1.16, in relation to HIST with 0.10, for example.

- *Cluster Analysis of the JCR categories*

Table 2 shows a Cluster Analysis of the JCR categories according to the AIF components. Ward's method is the criterion applied in the hierarchical cluster analysis. We consider two levels of aggregation. The first cluster level (L1) is configured by closest categories in the publication and citation habits. The second level (L2) contains meta-clusters of categories that are relatively closer in the publication and citation habits. Although some clusters exclusively contain categories from the same edition (C4 and C8), in most cases there are categories from both editions of the JCR. There are two very large clusters (C5 and C6), with more than 25% of categories each. Note that around 70% of categories are clustered in C12 while about 4% of categories are not clustered. The non clustered group includes some popular categories such as S17 (BIOL), S113 (MULTIDISCIP SCI), and SS20 (HIST).

[Table 2 about here]

- *Principal Component Analysis of the AIF*

Table 3 shows the correlation between the main variables. Note that increasing the average number of references in a field $r_t^F$ also increases the citations received $NCited_t^F$; correlations of 0.95 and 0.89 in Science and Social Science, respectively. Something similar occurs with the number of references to JCR items $J_t^F$. Moreover, if a field produces more citations $NCiting_t^F$, then it also receives more citations $NCited_t^F$; correlations of 0.98 and 0.87 in Science and Social Science, respectively.

[Table 3 about here]

The correlation between components in Table 3 is low or non-existent. There is a high correlation in Social Science between the ratio of references to JCR items $J_t^F$ and the proportion of cited to citing items $b_t^F$. In this edition, categories citing more JCR items (closer to categories in Science) receive more citations from outside the category.

In general, the components of the AIF correlate little or nothing with the AIF. This is shown in Figure 2. Whereas the correlation with the proportion cited to citing items is similar in both editions, the correlation with the ratio of references to JCR items is much higher in Social Science. The correlation with the average number of references is low in Science and there is no correlation in Social Science.

[Figure 2 about here]

Table 3 also shows the eigenvalues of a Principal Component Analysis (PCA). This eigenvalue decomposition of the correlation matrix in terms of component scores is used to find the causes of the variability in the dataset and sort them by importance. The principal components are guaranteed to be independent because the variables are normally distributed according to a Kolmogorov-Smirnov test (see also Table 4 and Figure 3).

[Table 4 and Figure 3 about here]

The analysis reveals that in Science the variance can be explained to a great degree by three major components: the ratio of references to JCR items ($J_t^F$, 36.55%), the ratio of JCR references to the target window ($w_t^F$, 20.93%), and the field growth ($a_t^F$, 20.60%). These components together explain 78.08% of the total variance. In Social Science, the variance can be explained to a great degree by only two major components: the ratio of JCR references to the target window ($w_t^F$, 57.79%) and the proportion of cited to citing items in the target window ($b_t^F$, 23.50%). These components together explain 81.29 % of the total variance.

- *Comparing IF and CNIF*

The CNIF is obtained for economics and business field journals in more than one JCR subject category (SS3, SS4, SS9, SS29, and S124). In each category, the *percentile* of a journal in the ranking is obtained as the position in the ranking divided by the number of journals in the category (*x100%*). A percentile is the value below which a certain percent of journals fall. The *gap* is defined as the maximum distance among the ranking percentiles from all categories where each journal is included.

For instance, COMPUT ECON is in percentiles 67 and 85 for its two categories according to IF. The same procedure is applied to ranking percentiles obtained with CNIF. COMPUT ECON is ranked in percentiles 69 and 77 for its two categories according to CNIF. Then, the gap according to IF is 18 (85-67) and according to CNIF is 8 (77-69).

There are important differences between the CNIF and the IF in most of the journals analyzed. The maximum gap in the IF case is 28 while in the CNIF case it is 17. The average gap in the IF case is 6.2 whereas the CNIF case is 4.2. Therefore, the maximum gap is reduced by around 39% and the average gap by about 32%. Moreover, in 51% of the journals analyzed the maximum gap has decreased.

Finally, Figures 4 and 5 show the effect of the normalization on the Impact Factor. In all cases the impact of the journals analyzed has increased, some more than others (in a quantity higher than two in some cases). Therefore, the economics and business field is being penalized by the Impact Factor in comparison to the rest of fields in the JCR.

[Figures 4 and 5 about here]

*4.3 Discussion*

The AIF in Science is around 58% higher than in Social Science. This is due to the fact that although on average there are over 30% more references in articles of Social Science, an important part of them are non-JCR items. In Social Science around 40% of the references are books and journals that are not indexed in the JCR, while in Science these are around 20% of the references.

There exists a great variance in the AIF within each edition, in some cases between journals of relatively close categories. For example, the AIF of category MATH & COMP BIOL (3.038) is almost four times greater than category MATH (0.829). The categories with the highest AIF in Science are related to biomedicine. The lowest values are in engineering and mathematics. In Social Science, the categories with the highest values are related to psychology and some specialties of economics, such as health policy and management. The lowest factors are in categories related to history.

The JCR database grows annually around 9%. The Social Science edition is growing over twice as much as the Science edition (16% and 7%, respectively). This is due to the incorporation of journals in some categories of the Social Science in the last few years.

A journal from a category of Social Science has on average 30% more references and around 20% more references to non-JCR items than a journal from a Science category. The longest reference lists are produced in history and the shortest in engineering and mathematics. The highest proportions to non-JCR items are in physics, biology, and chemistry, whereas the lowest are in engineering and computer science. In Social Science, the highest values are in psychology and the lowest in history.

One of each five JCR references is on average in the target window. Curiously, some of the categories with the lowest proportion of references to JCR items have the highest proportion of citations in the target window. This is due to the fact that the older references correspond to books while the newer references correspond to articles. This happens, for example, in engineering and history. In some areas, such as mathematics, just one in eight JCR references is from the previous two years, in comparison to history where they are one in three.

In general, the highest proportions cited to citing are in biomedicine and lowest are obtained in history and law. However, notice the exceptional case of category MULTIDISCIP SCI where more than half of the citations came from outside the category. In Social Science, categories citing more JCR items (closer to categories in Science) receive more citations from outside the category, some of them from Science categories.

With respect to the Cluster Analysis of the JCR categories, C1 and C7 include, in general, those life sciences with an important social component, as well as those social sciences which use mathematical methods in a higher degree (health, psychology, economics, and business, for example). However, there exist important differences between C1 and C7, and they are not clustered jointly in the second level. Notice that S119 and SS30 (NURS) are in the same cluster (C1), as there are not enough differences between them to justify the existence of two similar categories in Science and Social Science editions. A similar case occurs with S79 and SS21 (HIST & PHILOS OF SCI) in cluster C2. Note that ECON is in C1, however BUSINESS and MANAGE are in C7; this highlights the heterogeneity of the economics and business field (observe also that BUSINESS, FINANCE is in C6).

Clusters C2 and C4 contain those social sciences which use mathematical methods in a lower degree (education, sociology, linguistic, and law, for example). Finally, clusters C5 and C6 include formal, physical, technological, and life sciences (mathematics, physics, chemistry, engineering, and biomedicine, for example).

The differences across categories within the same edition are in some cases greater than those which exist between some categories from different editions. This is the case of GERONTOL and PSYCHIATRY, close to Science, and S79 (HIST & PHILOS OF SCI), close to Social Science, for example.

The variance in the AIF in Science can be explained to a great degree by three major components (the ratio of references to JCR items, the ratio of JCR references to the target window, and the field growth). In Social Science, the variance can be explained to a great degree by only two major components (the ratio of JCR references to the target window and the proportion of cited to citing items in the target window). The principal components are different depending on the edition of the JCR. This is motivated because in Social Science there are many different disciplines in relation to the habits of publication and citation (economics and psychology versus history, for example).

There are important differences between the CNIF and the IF for most of the journals analyzed. In the case of the CNIF, the maximum gap is reduced in more than half of the journals with respect to the IF. The average gap is also reduced by around a 32%.

**5. Conclusions**

The journal Impact Factor is not comparable among fields of science because of systematic differences in publication and citation behaviour across disciplines. A decomposing of the field aggregate impact factor into five normally distributed variables shows that, for the *JCR subject categories* of Science and Social Science, the variables that to a greater degree explain the variance in the impact factor of a field do not include the average number of references. However, this is the factor that has most frequently been used in the literature to justify the differences between fields of science, as well as the most employed in the source-normalization (Leydesdorff & Bornmann, 2011; Moed, 2010; Zitt & Small, 2008). Therefore, it is necessary to consider some other sources of variance in the normalization process.

In this sense, a normalized impact indicator based on the JCR subject category list is proposed and compared with the Impact Factor. This normalization is achieved by considering all the categories indexing each journal. An empirical application, with one hundred journals in two or more subject categories of economics and business, shows that the gap between rankings diminishes in one half of the cases analyzed and by around 32%.

The field considered (economics and business) is an example of a heterogeneous area that is penalized by the Impact Factor, but not the only one, since the normalized impact has increased in all journals analyzed. Additionally, there are some other fields that are favored by the Impact Factor. This is the main reason why it is necessary to be cautious when comparing journal impact factors from different fields. In this sense, our index has behaved well in a big number of journals.

Finally, we would emphasize the nature of JCR subject category data and the doubtfulness of obtaining results precise enough for the purpose no matter how sophisticated the mathematical and statistical technique.

**Acknowledgements**

This research has been supported by the Ministry of Science and Technology of Spain under the research project ECO2008-05589.

Table 1: Aggregate Impact Factor components of the JCR subject categories in the 2010 Science and Social Science editions (S=Science, SS=Social Science, *t*=2010, - Data not available)

| Code | JCR subject category (F) | References | | Citations to JCR items in the target window | | AIF components | | | | | Aggregate Impact Factor |
|---|---|---|---|---|---|---|---|---|---|---|---|
| | | $JCR\ J_t^F$ | $Total\ R_t^F$ | $NCited_t^F$ | $NCiting_t^F$ | $a_t^F$ | $r_t^F$ | $p_t^F$ | $w_t^F$ | $b_t^F$ | $AIF_t^F$ |
| S1 | ACOUSTICS | 87001 | 110560 | 11626 | 12872 | 0.51 | 29.14 | 0.79 | 0.15 | 0.90 | 1.553 |
| S2 | AGR ECON & POLICY | 10771 | 18343 | 1075 | 2395 | 0.48 | 38.94 | 0.59 | 0.22 | 0.45 | 1.088 |
| S3 | AGR ENGN | 65135 | 82827 | 13352 | 13182 | 0.65 | 29.92 | 0.79 | 0.20 | 1.01 | 3.123 |
| S4 | AGR, DAIRY & ANIMAL SCI | 165027 | 204241 | 15769 | 21064 | 0.55 | 33.81 | 0.81 | 0.13 | 0.75 | 1.428 |
| S5 | AGR, MULTIDISCIP | 140735 | 193124 | 15625 | 23783 | 0.63 | 32.96 | 0.73 | 0.17 | 0.66 | 1.673 |
| S6 | AGRONOMY | 191377 | 248439 | 19111 | 26277 | 0.62 | 37.44 | 0.77 | 0.14 | 0.73 | 1.774 |
| S7 | ALLERGY | 82572 | 94468 | 16195 | 18216 | 0.51 | 44.23 | 0.87 | 0.22 | 0.89 | 3.844 |
| S8 | ANATOMY & MORPHOL | 71567 | 85796 | 6011 | 9494 | 0.61 | 46.50 | 0.83 | 0.13 | 0.63 | 1.976 |
| S9 | ANDROL | 14097 | 16433 | 1571 | 2069 | 0.56 | 44.78 | 0.86 | 0.15 | 0.76 | 2.377 |
| S10 | ANESTHESIOL | 119952 | 136540 | 20567 | 24750 | 0.52 | 37.42 | 0.88 | 0.21 | 0.83 | 2.955 |
| S11 | ASTRON & ASTROPHYS | 587864 | 753996 | 131008 | 142737 | 0.47 | 56.59 | 0.78 | 0.24 | 0.92 | 4.609 |
| S12 | AUTOM & CONTROL SYST | 119510 | 171369 | 17595 | 23622 | 0.58 | 25.73 | 0.70 | 0.20 | 0.74 | 1.532 |
| S13 | BEHAV SCI | 277143 | 312477 | 32268 | 40224 | 0.51 | 57.76 | 0.89 | 0.15 | 0.80 | 3.048 |
| S14 | BIOCHEM RES METHODS | 497392 | 557611 | 98866 | 107235 | 0.56 | 38.37 | 0.89 | 0.22 | 0.92 | 3.822 |
| S15 | BIOCHEM & MOLEC BIOL | 2300858 | 2489594 | 428047 | 418919 | 0.52 | 49.62 | 0.92 | 0.18 | 1.02 | 4.435 |
| S16 | BIODIVERS CONSERVAT | 123090 | 159146 | 13946 | 19505 | 0.57 | 54.15 | 0.77 | 0.16 | 0.71 | 2.688 |
| S17 | BIOL | 653552 | 745525 | 74752 | 117221 | 0.85 | 48.06 | 0.88 | 0.18 | 0.64 | 4.114 |
| S18 | BIOPHYS | 467970 | 514619 | 76148 | 86868 | 0.51 | 43.37 | 0.91 | 0.19 | 0.88 | 3.291 |
| S19 | BIOTECH & APPL MICROBIOL | 822862 | 946753 | 137905 | 160035 | 0.57 | 39.23 | 0.87 | 0.19 | 0.86 | 3.256 |
| S20 | CARDIAC & CARDIO SYST | 554524 | 604904 | 115840 | 112024 | 0.59 | 37.98 | 0.92 | 0.20 | 1.03 | 4.277 |
| S21 | CELL & TISSUE ENGN | 75741 | 81945 | 18128 | 17421 | - | 75.66 | 0.92 | 0.23 | 1.04 | - |
| S22 | CELL BIOL | 1165303 | 1251961 | 268233 | 231292 | 0.54 | 55.27 | 0.93 | 0.20 | 1.16 | 6.453 |
| S23 | CHEM, ANALYT | 548695 | 628015 | 100590 | 119856 | 0.52 | 34.95 | 0.87 | 0.22 | 0.84 | 2.906 |
| S24 | CHEM, APPL | 317360 | 371130 | 49172 | 54563 | 0.52 | 31.82 | 0.86 | 0.17 | 0.90 | 2.207 |
| S25 | CHEM, INORG & NUCLEAR | 497463 | 538689 | 58993 | 79483 | 0.51 | 43.13 | 0.92 | 0.16 | 0.74 | 2.404 |
| S26 | CHEM, MED | 451855 | 516337 | 57426 | 88072 | 0.60 | 42.17 | 0.88 | 0.19 | 0.65 | 2.795 |
| S27 | CHEM, MULTIDISCIP | 1544644 | 1694928 | 324585 | 317737 | 0.59 | 40.37 | 0.91 | 0.21 | 1.02 | 4.586 |
| S28 | CHEM, ORGANIC | 765524 | 824033 | 111395 | 136404 | 0.51 | 41.38 | 0.93 | 0.18 | 0.82 | 2.853 |
| S29 | CHEM, PHYS | 1643567 | 1770821 | 285009 | 309709 | 0.57 | 39.72 | 0.93 | 0.19 | 0.92 | 3.615 |
| S30 | CLIN NEUROL | 779590 | 888013 | 132347 | 131593 | 0.56 | 38.73 | 0.88 | 0.17 | 1.01 | 3.238 |
| S31 | COMP SCI, ARTIF INTEL | 187274 | 285250 | 29613 | 33535 | 0.55 | 33.83 | 0.66 | 0.18 | 0.88 | 1.940 |
| S32 | COMP SCI, CYBERNET | 24119 | 41678 | 2843 | 4387 | 0.53 | 38.34 | 0.58 | 0.18 | 0.65 | 1.395 |
| S33 | COMP SCI, HARD & ARCHITEC | 54473 | 95160 | 8502 | 12437 | 0.51 | 26.30 | 0.57 | 0.23 | 0.68 | 1.203 |
| S34 | COMP SCI, INFORMAT SYST | 155262 | 266542 | 23113 | 36675 | 0.56 | 32.48 | 0.58 | 0.24 | 0.63 | 1.583 |
| S35 | COMP SCI, INTERDISCIP APPL | 227895 | 322163 | 30707 | 41310 | 0.53 | 32.46 | 0.71 | 0.18 | 0.74 | 1.652 |
| S36 | COMP SCI, SOFT ENGN | 106893 | 190047 | 14891 | 25990 | 0.53 | 29.61 | 0.56 | 0.24 | 0.57 | 1.240 |
| S37 | COMP SCI, THEORY & METHODS | 97412 | 164483 | 14231 | 18296 | 0.54 | 30.26 | 0.59 | 0.19 | 0.78 | 1.404 |
| S38 | CONSTRUCT & BUILD TECH | 57189 | 94782 | 7364 | 11475 | 0.59 | 24.49 | 0.60 | 0.20 | 0.64 | 1.121 |
| S39 | CRIT CARE MED | 153274 | 170072 | 29238 | 30200 | 0.54 | 42.06 | 0.90 | 0.20 | 0.97 | 3.924 |
| S40 | CRYSTALLOGRAPHY | 212892 | 231353 | 32921 | 41283 | 0.52 | 22.79 | 0.92 | 0.19 | 0.80 | 1.681 |
| S41 | DENTISTRY, ORAL SURG & MED | 218076 | 252356 | 26350 | 31184 | 0.55 | 34.00 | 0.86 | 0.14 | 0.84 | 1.966 |
| S42 | DERMATOL | 173103 | 201837 | 26506 | 30138 | 0.57 | 33.87 | 0.86 | 0.17 | 0.88 | 2.525 |
| S43 | DEV BIOL | 211739 | 229814 | 36751 | 36553 | 0.50 | 57.14 | 0.92 | 0.17 | 1.01 | 4.583 |
| S44 | ECOL | 640059 | 790606 | 84918 | 96279 | 0.53 | 54.00 | 0.81 | 0.15 | 0.88 | 3.094 |
| S45 | EDUC, SCI DISCIP | 48104 | 72082 | 5734 | 9403 | 0.67 | 28.67 | 0.67 | 0.20 | 0.61 | 1.529 |
| S46 | ELECTROCHEM | 302448 | 335975 | 56538 | 64829 | 0.67 | 31.88 | 0.90 | 0.21 | 0.87 | 3.615 |
| S47 | EMERGENCY MED | 68546 | 82846 | 8842 | 11669 | 0.66 | 30.25 | 0.83 | 0.17 | 0.76 | 2.123 |
| S48 | ENDOCRIN & METABOL | 611906 | 677374 | 113606 | 114554 | 0.55 | 46.80 | 0.90 | 0.19 | 0.99 | 4.304 |
| S49 | ENERGY & FUELS | 336244 | 437760 | 65702 | 84990 | 0.64 | 30.33 | 0.77 | 0.25 | 0.77 | 2.912 |
| S50 | ENGN, AEROSPACE | 30755 | 51921 | 3175 | 5620 | 0.43 | 23.96 | 0.59 | 0.18 | 0.56 | 0.628 |
| S51 | ENGN, BIOMED | 276851 | 328504 | 42038 | 49317 | 0.61 | 36.42 | 0.84 | 0.18 | 0.85 | 2.848 |
| S52 | ENGN, CHEM | 536301 | 653171 | 76256 | 96697 | 0.56 | 29.60 | 0.82 | 0.18 | 0.79 | 1.940 |
| S53 | ENGN, CIVIL | 199342 | 295089 | 30628 | 39031 | 0.57 | 27.06 | 0.68 | 0.20 | 0.78 | 1.593 |
| S54 | ENGN, ELECT & ELECTRON | 634985 | 876499 | 112849 | 137704 | 0.55 | 21.82 | 0.72 | 0.22 | 0.82 | 1.541 |
| S55 | ENGN, ENVIRONM | 259544 | 327262 | 52006 | 60964 | 0.58 | 35.44 | 0.79 | 0.23 | 0.85 | 3.258 |
| S56 | ENGN, GEOLOG | 36820 | 57045 | 3849 | 5303 | 0.55 | 30.52 | 0.65 | 0.14 | 0.73 | 1.132 |
| S57 | ENGN, INDUSTRIAL | 74614 | 105946 | 11104 | 11265 | 0.41 | 33.60 | 0.70 | 0.15 | 0.99 | 1.450 |
| S58 | ENGN, MANUFACT | 76839 | 107999 | 11339 | 12720 | 0.46 | 27.29 | 0.71 | 0.17 | 0.89 | 1.307 |
| S59 | ENGN, MARINE | 2376 | 6133 | 214 | 776 | 0.43 | 13.94 | 0.39 | 0.33 | 0.28 | 0.207 |
| S60 | ENGN, MECHAN | 210921 | 296658 | 25610 | 32354 | 0.54 | 24.35 | 0.71 | 0.15 | 0.79 | 1.127 |
| S61 | ENGN, MULTIDISCIP | 132667 | 199242 | 14276 | 23749 | 0.52 | 25.06 | 0.67 | 0.18 | 0.60 | 0.928 |
| S62 | ENGN, OCEAN | 13260 | 20338 | 1785 | 2306 | 0.48 | 23.59 | 0.65 | 0.17 | 0.77 | 0.998 |
| S63 | ENGN, PETROLEUM | 24595 | 36848 | 1577 | 3498 | 0.60 | 21.95 | 0.67 | 0.14 | 0.45 | 0.565 |
| S64 | ENTOMOL | 160936 | 213373 | 14138 | 21061 | 0.55 | 38.68 | 0.75 | 0.13 | 0.67 | 1.409 |
| S65 | ENVIRONM SCI | 863759 | 1135683 | 133091 | 169848 | 0.51 | 41.60 | 0.76 | 0.20 | 0.78 | 2.507 |
| S66 | EVOLUT BIOL | 268481 | 315846 | 40368 | 42326 | 0.53 | 60.69 | 0.85 | 0.16 | 0.95 | 4.116 |
| S67 | FISHERIES | 150855 | 190662 | 12664 | 18730 | 0.54 | 44.07 | 0.79 | 0.12 | 0.68 | 1.579 |
| S68 | FOOD SCI & TECH | 492829 | 604057 | 58996 | 76620 | 0.57 | 34.60 | 0.82 | 0.16 | 0.77 | 1.942 |

| | | | | | | | | | | |
|---|---|---:|---:|---:|---:|---:|---:|---:|---:|---:|
| S69 | FORESTRY | 118549 | 163891 | 10745 | 16473 | 0.54 | 45.07 | 0.72 | 0.14 | 0.65 | 1.607 |
| S70 | GASTROEN & HEPATOL | 382940 | 420138 | 77128 | 75259 | 0.52 | 40.17 | 0.91 | 0.20 | 1.02 | 3.801 |
| S71 | GENET & HERED | 759905 | 849856 | 156577 | 150409 | 0.53 | 49.95 | 0.89 | 0.20 | 1.04 | 4.861 |
| S72 | GEOCHEM & GEOPHYS | 324472 | 382922 | 33619 | 42544 | 0.54 | 49.78 | 0.85 | 0.13 | 0.79 | 2.358 |
| S73 | GEOGRAPHY, PHYS | 160804 | 210478 | 14939 | 22751 | 0.55 | 59.09 | 0.76 | 0.14 | 0.66 | 2.323 |
| S74 | GEOL | 93025 | 123641 | 7929 | 10439 | 0.51 | 57.08 | 0.75 | 0.11 | 0.76 | 1.868 |
| S75 | GEOSCI, MULTIDISCIP | 697181 | 873861 | 76510 | 95789 | 0.53 | 48.47 | 0.80 | 0.14 | 0.80 | 2.230 |
| S76 | GERIATR & GERONTOL | 138082 | 165247 | 19901 | 26049 | 0.56 | 46.67 | 0.84 | 0.19 | 0.76 | 3.158 |
| S77 | HEALTH CARE SCI & SERV | 151209 | 216231 | 21434 | 33354 | 0.62 | 35.22 | 0.70 | 0.22 | 0.64 | 2.154 |
| S78 | HEMATOL | 436060 | 472192 | 109358 | 92940 | 0.51 | 44.98 | 0.92 | 0.21 | 1.18 | 5.310 |
| S79 | HIST & PHILOS OF SCI | 27945 | 62939 | 1678 | 5515 | 0.59 | 48.08 | 0.44 | 0.20 | 0.30 | 0.754 |
| S80 | HORTICULTURE | 83424 | 105471 | 7471 | 11143 | 0.58 | 34.99 | 0.79 | 0.13 | 0.67 | 1.429 |
| S81 | IMAG SCI & PHOTO TECH | 45101 | 61322 | 5867 | 7468 | 0.60 | 37.78 | 0.74 | 0.17 | 0.79 | 2.186 |
| S82 | IMMUNOL | 826396 | 902431 | 176306 | 172838 | 0.51 | 45.73 | 0.92 | 0.21 | 1.02 | 4.585 |
| S83 | INFECTIOUS DIS | 296452 | 340308 | 67287 | 69288 | 0.54 | 36.20 | 0.87 | 0.23 | 0.97 | 3.879 |
| S84 | INSTRUM & INSTRUMENTAT | 205016 | 265740 | 35387 | 45316 | 0.54 | 23.50 | 0.77 | 0.22 | 0.78 | 1.675 |
| S85 | INTEGRAT & COMPL MED | 45667 | 61939 | 5667 | 8647 | 0.68 | 38.74 | 0.74 | 0.19 | 0.66 | 2.402 |
| S86 | LIMNOL | 73679 | 90960 | 6712 | 15236 | 0.59 | 46.86 | 0.81 | 0.21 | 0.44 | 2.028 |
| S87 | MARINE & FRESH BIOL | 353748 | 439034 | 33305 | 45527 | 0.50 | 49.14 | 0.81 | 0.13 | 0.73 | 1.870 |
| S88 | MAT SCI, BIOMAT | 161212 | 180555 | 27166 | 28420 | 0.63 | 39.30 | 0.89 | 0.18 | 0.96 | 3.729 |
| S89 | MAT SCI, CERAM | 80794 | 94437 | 10855 | 11926 | 0.45 | 24.40 | 0.86 | 0.15 | 0.91 | 1.264 |
| S90 | MAT SCI, CHARAC & TEST | 25396 | 38432 | 3345 | 4321 | 0.54 | 19.84 | 0.66 | 0.17 | 0.77 | 0.939 |
| S91 | MAT SCI, COAT & FILMS | 147205 | 162025 | 22364 | 24633 | 0.51 | 27.61 | 0.91 | 0.17 | 0.91 | 1.943 |
| S92 | MAT SCI, COMPOSITES | 52688 | 67083 | 6749 | 7675 | 0.58 | 26.58 | 0.79 | 0.15 | 0.88 | 1.553 |
| S93 | MAT SCI, MULTIDISCIP | 1534898 | 1718357 | 287829 | 309530 | 0.55 | 32.07 | 0.89 | 0.20 | 0.93 | 2.949 |
| S94 | MAT SCI, PAPER & WOOD | 21449 | 31552 | 2098 | 3852 | 0.56 | 24.59 | 0.68 | 0.18 | 0.54 | 0.912 |
| S95 | MAT SCI, TEXT | 23909 | 34522 | 3280 | 4267 | 0.55 | 23.01 | 0.69 | 0.18 | 0.77 | 1.208 |
| S96 | MATH & COMP BIOL | 148002 | 178634 | 23779 | 28791 | 0.61 | 37.38 | 0.83 | 0.19 | 0.83 | 3.038 |
| S97 | MATH | 304735 | 410885 | 30156 | 37579 | 0.55 | 20.49 | 0.74 | 0.12 | 0.80 | 0.829 |
| S98 | MATH, APPL | 371885 | 497222 | 43713 | 55245 | 0.60 | 23.68 | 0.75 | 0.15 | 0.79 | 1.247 |
| S99 | MATH, INTERDISCIP APPL | 154935 | 207574 | 19456 | 23504 | 0.52 | 30.90 | 0.75 | 0.15 | 0.83 | 1.515 |
| S100 | MECHAN | 325669 | 416158 | 40233 | 46736 | 0.56 | 28.83 | 0.78 | 0.14 | 0.86 | 1.574 |
| S101 | MED ETH | 12392 | 21839 | 1756 | 3911 | 0.56 | 35.05 | 0.57 | 0.32 | 0.45 | 1.581 |
| S102 | MED INFORMAT | 43039 | 63614 | 6730 | 9045 | 0.56 | 31.98 | 0.68 | 0.21 | 0.74 | 1.893 |
| S103 | MED LAB TECH | 84687 | 100724 | 11600 | 16603 | 0.55 | 34.80 | 0.84 | 0.20 | 0.70 | 2.208 |
| S104 | MED, GEN & INTERNAL | 564352 | 707658 | 145040 | 123869 | 0.61 | 38.13 | 0.80 | 0.22 | 1.17 | 4.754 |
| S105 | MED, LEGAL | 31589 | 45286 | 4010 | 8185 | 0.59 | 34.18 | 0.70 | 0.26 | 0.49 | 1.787 |
| S106 | MED, RES & EXPT | 558834 | 632036 | 84168 | 111221 | 0.61 | 46.04 | 0.88 | 0.20 | 0.76 | 3.753 |
| S107 | METALL & METALL ENGN | 256599 | 309126 | 33430 | 45099 | 0.52 | 23.94 | 0.83 | 0.18 | 0.74 | 1.346 |
| S108 | METEOROL & ATMOS SCI | 279224 | 336111 | 37740 | 46993 | 0.55 | 40.13 | 0.83 | 0.17 | 0.80 | 2.475 |
| S109 | MICROBIOL | 656715 | 729125 | 119330 | 120871 | 0.55 | 42.59 | 0.90 | 0.18 | 0.99 | 3.801 |
| S110 | MICROSCOPY | 27154 | 33809 | 4693 | 4851 | 0.48 | 34.15 | 0.80 | 0.18 | 0.97 | 2.293 |
| S111 | MINERAL | 78015 | 95651 | 7341 | 9567 | 0.51 | 45.29 | 0.82 | 0.12 | 0.77 | 1.790 |
| S112 | MIN & MINERAL PROC | 27955 | 40715 | 4039 | 4725 | 0.46 | 22.67 | 0.69 | 0.17 | 0.85 | 1.033 |
| S113 | MULTIDISCIP SCI | 379543 | 451475 | 206138 | 80965 | 0.58 | 36.81 | 0.84 | 0.21 | 2.55 | 9.707 |
| S114 | MYCOL | 53821 | 66081 | 5867 | 8038 | 0.58 | 39.69 | 0.81 | 0.15 | 0.73 | 2.059 |
| S115 | NANOSCI & NANOTECH | 661585 | 722178 | 145992 | 152401 | 0.60 | 35.80 | 0.92 | 0.23 | 0.96 | 4.365 |
| S116 | NEUROIMAG | 93439 | 101390 | 15479 | 15648 | 0.57 | 47.16 | 0.92 | 0.17 | 0.99 | 4.098 |
| S117 | NEUROSCI | 1618720 | 1768206 | 245526 | 259773 | 0.53 | 55.03 | 0.92 | 0.16 | 0.95 | 4.082 |
| S118 | NUCLEAR SCI & TECH | 113692 | 151346 | 16609 | 21096 | 0.49 | 19.21 | 0.75 | 0.19 | 0.79 | 1.025 |
| S119 | NURS | 125366 | 191459 | 10869 | 26587 | 0.66 | 36.50 | 0.65 | 0.21 | 0.41 | 1.369 |
| S120 | NUTRIT & DIETET | 286862 | 342612 | 46423 | 49255 | 0.54 | 42.05 | 0.84 | 0.17 | 0.94 | 3.098 |
| S121 | OBSTETR & GYNECOL | 301783 | 350442 | 44212 | 54848 | 0.56 | 33.92 | 0.86 | 0.18 | 0.81 | 2.397 |
| S122 | OCEANOGRAPHY | 192373 | 238310 | 18538 | 24957 | 0.53 | 47.04 | 0.81 | 0.13 | 0.74 | 1.943 |
| S123 | ONCOL | 1061251 | 1162357 | 248653 | 229993 | 0.55 | 41.65 | 0.91 | 0.22 | 1.08 | 4.941 |
| S124 | OPER RES & MANAGE SCI | 153260 | 212548 | 20324 | 21253 | 0.52 | 31.46 | 0.72 | 0.14 | 0.96 | 1.557 |
| S125 | OPHTHALMOL | 241331 | 268410 | 34700 | 38015 | 0.52 | 35.06 | 0.90 | 0.16 | 0.91 | 2.379 |
| S126 | OPTICS | 458808 | 516733 | 83455 | 95436 | 0.56 | 24.26 | 0.89 | 0.21 | 0.87 | 2.204 |
| S127 | ORNITHOL | 38828 | 50439 | 2483 | 4803 | 0.51 | 47.14 | 0.77 | 0.12 | 0.52 | 1.182 |
| S128 | ORTHOPED | 243535 | 274007 | 31353 | 32722 | 0.57 | 31.38 | 0.89 | 0.13 | 0.96 | 2.048 |
| S129 | OTORHIN | 107942 | 128555 | 12514 | 15689 | 0.60 | 25.91 | 0.84 | 0.15 | 0.80 | 1.501 |
| S130 | PALEONTOL | 114685 | 155830 | 7777 | 11907 | 0.56 | 67.05 | 0.74 | 0.10 | 0.65 | 1.873 |
| S131 | PARASITOL | 158367 | 184377 | 23053 | 29583 | 0.58 | 42.14 | 0.86 | 0.19 | 0.78 | 3.056 |
| S132 | PATHOL | 253772 | 288633 | 36577 | 47847 | 0.57 | 38.58 | 0.88 | 0.19 | 0.76 | 2.763 |
| S133 | PEDIATR | 359622 | 430521 | 49629 | 63818 | 0.55 | 31.37 | 0.84 | 0.18 | 0.78 | 2.005 |
| S134 | PERIPH VASCULAR DIS | 355361 | 387120 | 79062 | 70290 | 0.56 | 40.40 | 0.92 | 0.20 | 1.12 | 4.612 |
| S135 | PHARMACOL & PHARM | 1316587 | 1505438 | 184677 | 268016 | 0.54 | 47.69 | 0.87 | 0.20 | 0.69 | 3.134 |
| S136 | PHYS, APPL | 945463 | 1053047 | 215235 | 203512 | 0.52 | 25.40 | 0.90 | 0.22 | 1.06 | 2.724 |
| S137 | PHYS, ATOM, MOLEC & CHEM | 571932 | 609551 | 69842 | 91796 | 0.50 | 40.57 | 0.94 | 0.16 | 0.76 | 2.344 |
| S138 | PHYS, CONDEN MATTER | 783797 | 842466 | 156603 | 151310 | 0.53 | 31.50 | 0.93 | 0.19 | 1.03 | 3.095 |
| S139 | PHYS, FLUIDS & PLASM | 214211 | 240795 | 29475 | 34286 | 0.57 | 30.99 | 0.89 | 0.16 | 0.86 | 2.151 |
| S140 | PHYS, MATH | 261378 | 309310 | 35684 | 41979 | 0.49 | 30.84 | 0.85 | 0.16 | 0.85 | 1.726 |
| S141 | PHYS, MULTIDISCIP | 573143 | 653806 | 133654 | 111266 | 0.49 | 30.25 | 0.88 | 0.19 | 1.20 | 3.046 |
| S142 | PHYS, NUCLEAR | 161221 | 183704 | 19947 | 26941 | 0.52 | 31.99 | 0.88 | 0.17 | 0.74 | 1.796 |
| S143 | PHYS, PARTIC & FIELDS | 371309 | 411286 | 67830 | 80556 | 0.53 | 40.39 | 0.90 | 0.22 | 0.84 | 3.503 |
| S144 | PHYSIOL | 460753 | 508333 | 62539 | 69011 | 0.51 | 51.08 | 0.91 | 0.15 | 0.91 | 3.223 |
| S145 | PLANT SCI | 672802 | 801562 | 84660 | 98536 | 0.56 | 45.81 | 0.84 | 0.15 | 0.86 | 2.692 |
| S146 | POLYMER SCI | 507387 | 564922 | 70658 | 85871 | 0.55 | 36.67 | 0.90 | 0.17 | 0.82 | 2.508 |
| S147 | PRIMARY HEALTH CARE | 26602 | 37017 | 3279 | 6877 | - | 35.46 | 0.72 | 0.26 | 0.48 | - |

| Code | Field | | | | | | | | | |
|---|---|---|---|---|---|---|---|---|---|---|
| S148 | PSYCHIATRY | 485603 | 576782 | 79435 | 81974 | 0.54 | 47.13 | 0.84 | 0.17 | 0.97 | 3.507 |
| S149 | PSYCHOL | 211860 | 257868 | 25390 | 28636 | 0.54 | 51.79 | 0.82 | 0.14 | 0.89 | 2.741 |
| S150 | PUBLIC, ENVIRONM & OCC GEN HEALTH | 385850 | 519875 | 65623 | 83835 | 0.60 | 35.30 | 0.74 | 0.22 | 0.78 | 2.666 |
| S151 | RADIOL, NUCL MED & MED IMAG | 481649 | 545932 | 85883 | 91399 | 0.56 | 34.00 | 0.88 | 0.19 | 0.94 | 2.972 |
| S152 | REHABILITAT | 99117 | 124731 | 10646 | 15214 | 0.65 | 38.16 | 0.79 | 0.15 | 0.70 | 2.103 |
| S153 | REMOTE SENS | 51285 | 69054 | 7321 | 9266 | 0.55 | 33.12 | 0.74 | 0.18 | 0.79 | 1.948 |
| S154 | REPRODUCTIVE BIOL | 179750 | 198502 | 23610 | 26713 | 0.54 | 44.88 | 0.91 | 0.15 | 0.88 | 2.904 |
| S155 | RESPIRATORY SYST | 240911 | 268308 | 45504 | 46199 | 0.53 | 38.69 | 0.90 | 0.19 | 0.98 | 3.475 |
| S156 | RHEUMATOL | 157097 | 174265 | 32264 | 31641 | 0.56 | 39.81 | 0.90 | 0.20 | 1.02 | 4.133 |
| S157 | ROBOT | 20108 | 32676 | 3355 | 4136 | 0.60 | 29.23 | 0.62 | 0.21 | 0.81 | 1.795 |
| S158 | SOIL SCI | 130188 | 161955 | 12098 | 15840 | 0.52 | 44.37 | 0.80 | 0.12 | 0.76 | 1.721 |
| S159 | SPECTROSCOPY | 178154 | 208226 | 26031 | 32306 | 0.50 | 32.72 | 0.86 | 0.18 | 0.81 | 2.065 |
| S160 | SPORT SCI | 222816 | 264736 | 28069 | 31732 | 0.58 | 37.59 | 0.84 | 0.14 | 0.88 | 2.300 |
| S161 | STAT & PROBABIL | 138380 | 189435 | 16634 | 18831 | 0.53 | 26.86 | 0.73 | 0.14 | 0.88 | 1.241 |
| S162 | SUBSTANCE ABUSE | 60317 | 73043 | 8049 | 9739 | 0.54 | 49.29 | 0.83 | 0.16 | 0.83 | 2.959 |
| S163 | SURGERY | 747985 | 844558 | 120918 | 123438 | 0.56 | 28.51 | 0.89 | 0.17 | 0.98 | 2.272 |
| S164 | TELECOM | 124425 | 197889 | 21791 | 31501 | 0.55 | 21.95 | 0.63 | 0.25 | 0.69 | 1.331 |
| S165 | THERMODYN | 132560 | 171792 | 17614 | 21357 | 0.56 | 27.94 | 0.77 | 0.16 | 0.82 | 1.608 |
| S166 | TOXICOL | 362715 | 433917 | 47847 | 62011 | 0.54 | 46.25 | 0.84 | 0.17 | 0.77 | 2.765 |
| S167 | TRANSPLANT | 134825 | 148395 | 27423 | 27311 | 0.51 | 30.50 | 0.91 | 0.20 | 1.00 | 2.876 |
| S168 | TRANSPORT SCI & TECH | 38432 | 64196 | 4341 | 9395 | 0.60 | 23.77 | 0.60 | 0.24 | 0.46 | 0.957 |
| S169 | TROP MED | 75960 | 94259 | 10570 | 15699 | 0.64 | 33.26 | 0.81 | 0.21 | 0.67 | 2.400 |
| S170 | UROL & NEPHROL | 305917 | 339449 | 58233 | 65473 | 0.51 | 35.21 | 0.90 | 0.21 | 0.89 | 3.078 |
| S171 | VETERINARY SCI | 354658 | 447730 | 31900 | 49663 | 0.53 | 32.16 | 0.79 | 0.14 | 0.64 | 1.213 |
| S172 | VIROL | 287583 | 307702 | 47242 | 56952 | 0.56 | 48.09 | 0.93 | 0.20 | 0.83 | 4.122 |
| S173 | WATER RESOURCES | 246190 | 337415 | 30246 | 46250 | 0.55 | 35.51 | 0.73 | 0.19 | 0.65 | 1.764 |
| S174 | ZOOL | 361803 | 460891 | 30040 | 44346 | 0.53 | 46.65 | 0.79 | 0.12 | 0.68 | 1.613 |
| SS1 | ANTHROPOL | 90985 | 160999 | 6675 | 16147 | 0.57 | 58.42 | 0.57 | 0.18 | 0.41 | 1.381 |
| SS2 | AREA STUDIES | 28124 | 81449 | 1537 | 12555 | 0.74 | 45.71 | 0.35 | 0.45 | 0.12 | 0.640 |
| SS3 | BUSINESS | 196820 | 276839 | 14674 | 24557 | 0.57 | 61.33 | 0.71 | 0.13 | 0.60 | 1.845 |
| SS4 | BUSINESS, FINANCE | 85960 | 117663 | 8184 | 13133 | 0.60 | 38.18 | 0.73 | 0.15 | 0.62 | 1.602 |
| SS5 | COMMUNICAT | 52751 | 98905 | 4104 | 11600 | 0.65 | 47.39 | 0.53 | 0.22 | 0.35 | 1.271 |
| SS6 | CRIMINOL & PENOL | 46888 | 80085 | 2842 | 7873 | 0.67 | 52.86 | 0.59 | 0.17 | 0.36 | 1.260 |
| SS7 | CULTURAL STUDIES | 4798 | 15095 | 268 | 1530 | - | 41.58 | 0.32 | 0.32 | 0.18 | - |
| SS8 | DEMOGRAPHY | 19467 | 36141 | 1680 | 3791 | 0.59 | 45.98 | 0.54 | 0.19 | 0.44 | 1.258 |
| SS9 | ECON | 324730 | 524600 | 32935 | 61834 | 0.64 | 36.42 | 0.62 | 0.19 | 0.53 | 1.459 |
| SS10 | EDUC & EDUC RES | 165628 | 310756 | 12141 | 32501 | 0.69 | 46.32 | 0.53 | 0.20 | 0.37 | 1.242 |
| SS11 | EDUC, SPECIAL | 36537 | 53139 | 2612 | 5600 | 0.66 | 48.80 | 0.69 | 0.15 | 0.47 | 1.574 |
| SS12 | ENVIRONM STUDIES | 124336 | 226111 | 13860 | 34232 | 0.65 | 50.81 | 0.55 | 0.28 | 0.40 | 2.027 |
| SS13 | ERGONOM | 26352 | 42035 | 2807 | 4380 | 0.53 | 40.77 | 0.63 | 0.17 | 0.64 | 1.436 |
| SS14 | ETHICS | 42523 | 71361 | 3487 | 9109 | 0.55 | 45.83 | 0.60 | 0.21 | 0.38 | 1.232 |
| SS15 | ETHNIC STUDIES | 11649 | 25591 | 813 | 2517 | 0.81 | 46.78 | 0.46 | 0.22 | 0.32 | 1.203 |
| SS16 | FAMILY STUDIES | 54465 | 84627 | 4028 | 8513 | 0.63 | 48.41 | 0.64 | 0.16 | 0.47 | 1.449 |
| SS17 | GEOGRAPHY | 79737 | 156234 | 7161 | 19868 | 0.61 | 58.80 | 0.51 | 0.25 | 0.36 | 1.644 |
| SS18 | GERONTOL | 66854 | 90104 | 8758 | 11599 | 0.54 | 44.13 | 0.74 | 0.17 | 0.76 | 2.335 |
| SS19 | HEALTH POLICY & SERV | 89450 | 135784 | 13659 | 21954 | 0.61 | 37.07 | 0.66 | 0.25 | 0.62 | 2.271 |
| SS20 | HIST | 19717 | 66282 | 612 | 6263 | 0.78 | 66.28 | 0.30 | 0.32 | 0.10 | 0.479 |
| SS21 | HIST & PHILOS OF SCI | 24325 | 52861 | 1540 | 4608 | 0.60 | 52.91 | 0.46 | 0.19 | 0.33 | 0.922 |
| SS22 | HIST OF SOCIAL SCI | 21007 | 50056 | 675 | 2684 | 0.71 | 65.09 | 0.42 | 0.13 | 0.25 | 0.623 |
| SS23 | HOSPITAL, LEIS, SPORT & TOUR | 48631 | 76963 | 2840 | 6860 | 0.97 | 61.92 | 0.63 | 0.14 | 0.41 | 2.212 |
| SS24 | INDUSTR RELAT & LABOR | 14563 | 27507 | 1076 | 2967 | 0.72 | 42.85 | 0.53 | 0.20 | 0.36 | 1.208 |
| SS25 | INFORMAT SCI & LIBR SCI | 63546 | 114676 | 7377 | 16483 | 0.57 | 38.89 | 0.55 | 0.26 | 0.45 | 1.430 |
| SS26 | INT RELAT | 56150 | 121969 | 4307 | 19684 | 0.63 | 48.10 | 0.46 | 0.35 | 0.22 | 1.078 |
| SS27 | LAW | 149174 | 230820 | 9451 | 41859 | 0.59 | 62.20 | 0.65 | 0.28 | 0.23 | 1.495 |
| SS28 | LINGUIST | 103509 | 180385 | 6186 | 15290 | 0.78 | 55.06 | 0.57 | 0.15 | 0.40 | 1.471 |
| SS29 | MANAGE | 257814 | 367462 | 20293 | 28323 | 0.64 | 63.55 | 0.70 | 0.11 | 0.72 | 2.249 |
| SS30 | NURS | 122147 | 187498 | 10510 | 25955 | 0.67 | 36.47 | 0.65 | 0.21 | 0.40 | 1.367 |
| SS31 | PLANN & DEV | 51099 | 102847 | 4417 | 12526 | 0.59 | 48.33 | 0.50 | 0.25 | 0.35 | 1.233 |
| SS32 | POLIT SCI | 108435 | 233178 | 8219 | 32227 | 0.62 | 46.49 | 0.47 | 0.30 | 0.26 | 1.011 |
| SS33 | PSYCHIATRY | 299136 | 374398 | 45075 | 50710 | 0.56 | 47.75 | 0.80 | 0.17 | 0.89 | 3.215 |
| SS34 | PSYCHOL, APPL | 100878 | 140896 | 8445 | 11509 | 0.53 | 57.51 | 0.72 | 0.11 | 0.73 | 1.812 |
| SS35 | PSYCHOL, BIOL | 59120 | 67642 | 5967 | 7775 | 0.53 | 57.76 | 0.87 | 0.13 | 0.77 | 2.682 |
| SS36 | PSYCHOL, CLIN | 208588 | 270475 | 24720 | 30799 | 0.55 | 49.03 | 0.77 | 0.15 | 0.80 | 2.459 |
| SS37 | PSYCHOL, DEV | 151501 | 195736 | 17088 | 19905 | 0.55 | 53.26 | 0.77 | 0.13 | 0.86 | 2.572 |
| SS38 | PSYCHOL, EDUC | 59230 | 87070 | 4686 | 7415 | 0.58 | 52.17 | 0.68 | 0.13 | 0.63 | 1.637 |
| SS39 | PSYCHOL, EXPT | 237748 | 288157 | 25895 | 30219 | 0.56 | 51.19 | 0.83 | 0.13 | 0.86 | 2.590 |
| SS40 | PSYCHOL, MATH | 13641 | 18429 | 2087 | 1792 | 0.49 | 33.51 | 0.74 | 0.13 | 1.16 | 1.840 |
| SS41 | PSYCHOL, MULTIDISCIP | 196309 | 282212 | 20743 | 29384 | 0.58 | 49.04 | 0.70 | 0.15 | 0.71 | 2.098 |
| SS42 | PSYCHOL, PSYCHOANAL | 11231 | 17756 | 976 | 1773 | 0.47 | 44.84 | 0.63 | 0.16 | 0.55 | 1.147 |
| SS43 | PSYCHOL, SOCIAL | 115913 | 155782 | 10135 | 14125 | 0.56 | 49.99 | 0.74 | 0.12 | 0.72 | 1.835 |
| SS44 | PUBLIC ADM | 35026 | 71439 | 2677 | 9244 | 0.64 | 50.31 | 0.49 | 0.26 | 0.29 | 1.199 |
| SS45 | PUBLIC, ENVIRONM & OCC GEN HEALTH | 242429 | 360171 | 30097 | 50332 | 0.66 | 39.61 | 0.67 | 0.21 | 0.60 | 2.177 |
| SS46 | REHABILITAT | 87309 | 123939 | 6987 | 13096 | 0.64 | 45.35 | 0.70 | 0.15 | 0.53 | 1.632 |
| SS47 | SOCIAL ISSUES | 24734 | 50252 | 2550 | 6875 | 0.55 | 37.14 | 0.49 | 0.28 | 0.37 | 1.043 |
| SS48 | SOCIAL SCI, BIOMED | 60793 | 92668 | 7862 | 12707 | 0.55 | 42.65 | 0.66 | 0.21 | 0.62 | 2.002 |
| SS49 | SOCIAL SCI, INTERDISCIP | 81449 | 156009 | 7010 | 16636 | 0.63 | 43.20 | 0.52 | 0.20 | 0.42 | 1.227 |
| SS50 | SOCIAL SCI, MATH METH | 43700 | 63868 | 4657 | 5666 | 0.57 | 33.54 | 0.68 | 0.13 | 0.82 | 1.392 |
| SS51 | SOCIAL WORK | 45449 | 81617 | 2937 | 8780 | 0.68 | 49.40 | 0.56 | 0.19 | 0.33 | 1.201 |
| SS52 | SOCIOL | 104512 | 222709 | 7693 | 22281 | 0.60 | 53.82 | 0.47 | 0.21 | 0.35 | 1.111 |

| | | | | | | | | | | |
|---|---|---|---|---|---|---|---|---|---|---|
| SS53 | SUBSTANCE ABUSE | 61928 | 83290 | 6596 | 10506 | 0.62 | 45.79 | 0.74 | 0.17 | 0.63 | 2.261 |
| SS54 | TRANSPORT | 26648 | 44866 | 3392 | 5717 | 0.68 | 36.42 | 0.59 | 0.21 | 0.59 | 1.874 |
| SS55 | URBAN STUDIES | 36971 | 71615 | 3003 | 7577 | 0.58 | 50.19 | 0.52 | 0.20 | 0.40 | 1.211 |
| SS56 | WOMEN'S STUDIES | 36072 | 65185 | 2381 | 6659 | 0.61 | 46.66 | 0.55 | 0.18 | 0.36 | 1.048 |
| *AIF of JCR (both editions)* | | | | | | | | | | | 2.822 |
| *AIF of Science edition* | | | | | | | | | | | 2.920 |
| *AIF of Social Science edition* | | | | | | | | | | | 1.848 |
| *JCR average* | | 261104 | 315665 | 42074 | 48080 | 0.57 | 39.88 | 0.75 | 0.18 | 0.74 | 2.258 |
| *JCR standard deviation* | | | | | | 0.07 | 10.69 | 0.14 | 0.05 | 0.25 | 1.183 |
| *Science average* | | 316816 | 372510 | 52894 | 58378 | 0.55 | 37.18 | 0.80 | 0.18 | 0.82 | 2.473 |
| *Science standard deviation* | | | | | | 0.05 | 10.01 | 0.10 | 0.04 | 0.21 | 1.248 |
| *Social Science average* | | 87999 | 139039 | 8453 | 16080 | 0.62 | 48.28 | 0.60 | 0.20 | 0.50 | 1.585 |
| *Social Science standard deviation* | | | | | | 0.09 | 8.09 | 0.13 | 0.07 | 0.22 | 0.562 |

Figure 1: Aggregate Impact Factor of the JCR subject categories (in decreasing order)

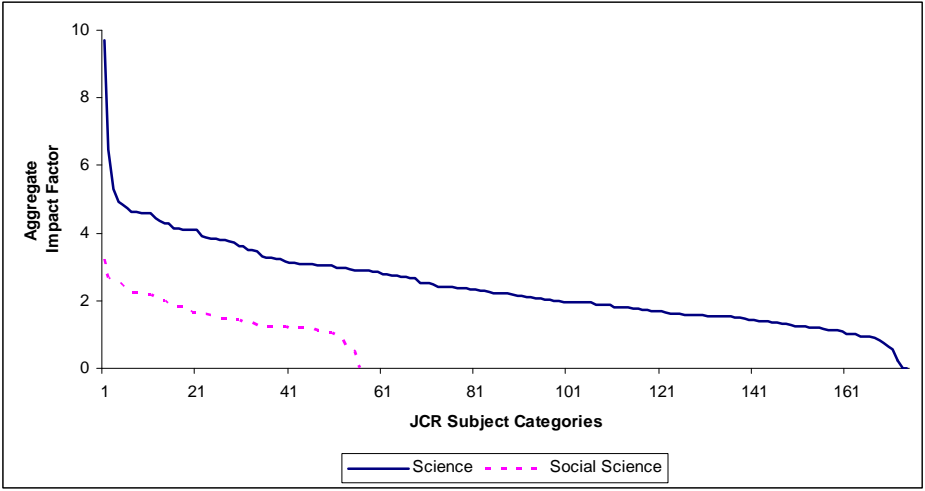

Table 2: Cluster Analysis of the JCR categories according to the AIF components

| Level | Cluster | # Categories | Science categories | Social Science categories |
|---|---|---|---|---|
| L1 | C1 | 29 (12.61%) | 2, 5, 32, 33, 34, 36, 38, 45, 50, 61, 77, 85, 86, 94, 103, 105, 119, 164, 168, 169, 173 | 9, 13, 19, 30, 42, 45, 48, 54 |
|  | C2 | 19 (8.26%) | 79 | 1, 5, 6, 8, 10, 11, 14, 15, 16, 21, 24, 28, 46, 49, 51, 52, 55, 56 |
|  | C3 | 3 (1.30%) | 101 | 25, 47 |
|  | C4 | 7 (3.04%) | - | 12, 17, 26, 27, 31, 32, 44 |
|  | C5 | 63 (27.39%) | 3, 7, 10, 14, 15, 18, 19, 20, 22, 23, 25, 26, 27, 28, 29, 30, 39, 46, 48, 49, 51, 55, 65, 70, 71, 76, 78, 82, 83, 88, 91, 93, 96, 104, 106, 108, 109, 110, 115, 116, 120, 123, 131, 132, 134, 135, 136, 137, 138, 141, 143, 148, 150, 151, 155, 156, 163, 166, 167, 170, 172 | 18, 33 |
|  | C6 | 59 (25.65%) | 1, 4, 6, 12, 24, 31, 35, 37, 40, 41, 42, 47, 52, 53, 54, 56, 57, 58, 60, 62, 64, 68, 80, 81, 84, 89, 90, 92, 95, 97, 98, 99, 100, 102, 112, 114, 118, 121, 124, 125, 126, 128, 129, 133, 139, 140, 142, 146, 152, 153, 157, 159, 160, 161, 165, 171 | 4, 40, 50 |
|  | C7 | 34 (14.78%) | 8, 9, 13, 16, 44, 67, 69, 72, 73, 74, 75, 87, 111, 122, 127, 130, 144, 145, 149, 154, 158, 162, 174 | 3, 29, 34, 35, 36, 37, 38, 39, 41, 43, 53 |
|  | C8 | 4 (1.74%) | 11, 43, 66, 117 | - |
|  | C9 | 3 (1.30%) | 107 | 22, 23 |
|  | Non clustered | 9 (3.91%) | 17, 21, 59, 63, 113, 147 | 2, 7, 20 |
| L2 | C10 | 48 (20.87%) | C1, C2 | |
|  | C11 | 10 (4.35%) | C3, C4 | |
|  | C12 | 160 (69.56%) | C5, C6, C7, C8 | |

Table 3: Correlations between the main variables and Principal Component Analysis (t=2010)

*Science subject categories*

| | /F/ | $A_t^F$ | $R_t^F$ | $J_t^F$ | $NCited_t^F$ | $NCiting_t^F$ | | $a_t^F$ | $r_t^F$ | $p_t^F$ | $w_t^F$ | $b_t^F$ | $AIF_t^F$ |
|---|---|---|---|---|---|---|---|---|---|---|---|---|---|
| /F/ | 1 | 0.81 | 0.79 | 0.75 | 0.67 | 0.72 | $a_t^F$ | 1 | 0.02 | 0.03 | 0.08 | -0.11 | 0.14 |
| $A_t^F$ | | 1 | 0.94 | 0.93 | 0.90 | 0.93 | $r_t^F$ | | 1 | 0.40 | -0.21 | 0.14 | 0.52 |
| $R_t^F$ | | | 1 | 1.00 | 0.95 | 0.98 | $p_t^F$ | | | 1 | -0.20 | 0.55 | 0.65 |
| $J_t^F$ | | | | 1 | 0.96 | 0.99 | $w_t^F$ | | | | 1 | -0.03 | 0.24 |
| $NCited_t^F$ | | | | | 1 | 0.98 | $b_t^F$ | | | | | 1 | 0.76 |
| $NCiting_t^F$ | | | | | | 1 | $AIF_t^F$ | | | | | | 1 |
| | | | | | | | PCA scores | 0.2060 | 0.0731 | 0.3655 | 0.2093 | 0.1460 | |

*Social Science subject categories*

| | /F/ | $A_t^F$ | $R_t^F$ | $J_t^F$ | $NCited_t^F$ | $NCiting_t^F$ | | $a_t^F$ | $r_t^F$ | $p_t^F$ | $w_t^F$ | $b_t^F$ | $AIF_t^F$ |
|---|---|---|---|---|---|---|---|---|---|---|---|---|---|
| /F/ | 1 | 0.90 | 0.90 | 0.80 | 0.66 | 0.85 | $a_t^F$ | 1 | 0.29 | -0.50 | 0.25 | -0.56 | -0.26 |
| $A_t^F$ | | 1 | 0.96 | 0.92 | 0.87 | 0.94 | $r_t^F$ | | 1 | -0.15 | -0.11 | -0.29 | -0.01 |
| $R_t^F$ | | | 1 | 0.97 | 0.89 | 0.95 | $p_t^F$ | | | 1 | -0.71 | 0.88 | 0.85 |
| $J_t^F$ | | | | 1 | 0.95 | 0.91 | $w_t^F$ | | | | 1 | -0.68 | -0.48 |
| $NCited_t^F$ | | | | | 1 | 0.87 | $b_t^F$ | | | | | 1 | 0.78 |
| $NCiting_t^F$ | | | | | | 1 | $AIF_t^F$ | | | | | | 1 |
| | | | | | | | PCA scores | 0.1173 | 0.0220 | 0.0478 | 0.5779 | 0.2350 | |

Figure 2: Scatter plot of the components with the Aggregate Impact Factor

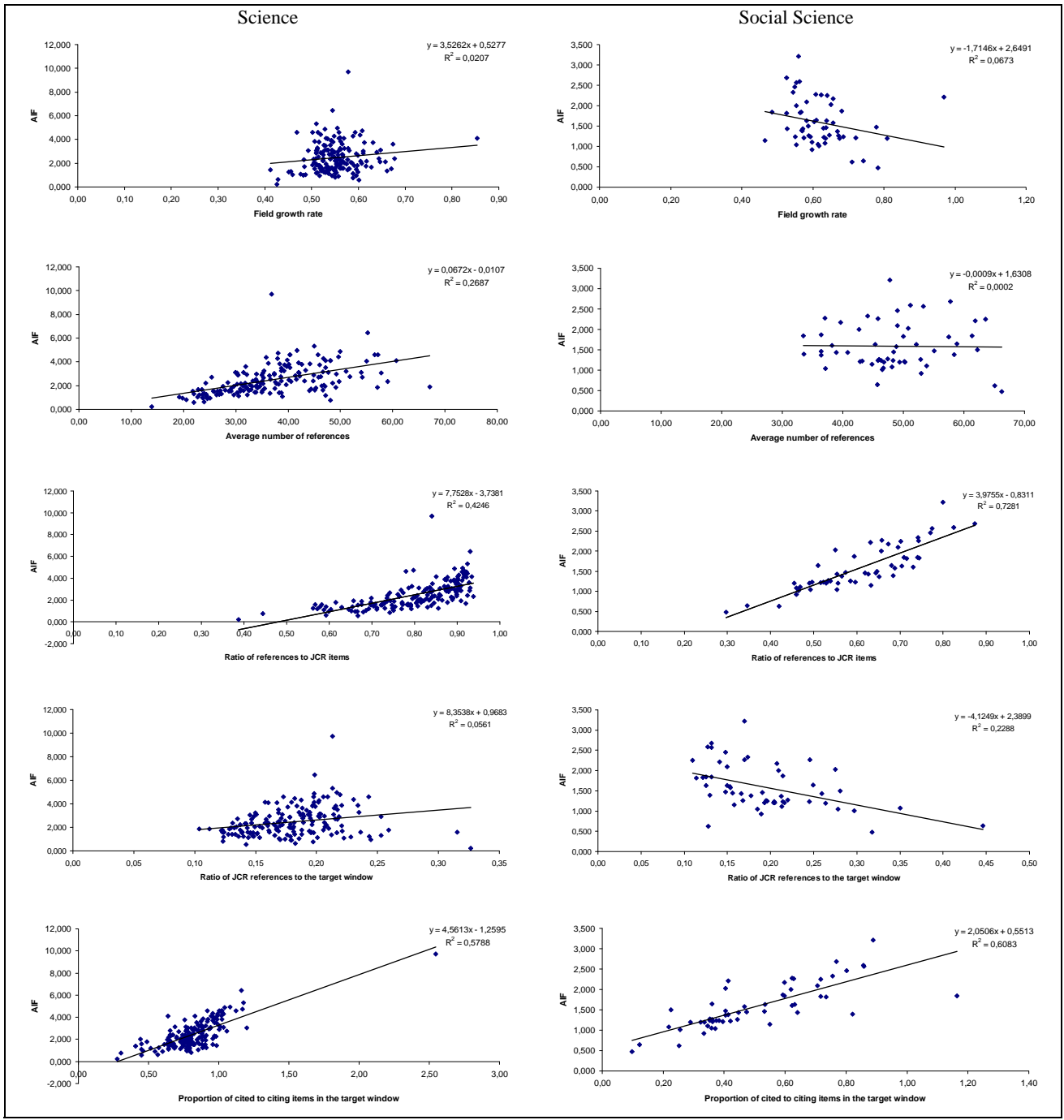

Table 4: Frequency of the components (*m*=mean, *s*=standard deviation)

| Interval | Science frequency | | | | | Social Science frequency | | | | |
| --- | --- | --- | --- | --- | --- | --- | --- | --- | --- | --- |
| | $a_t^F$ | $r_t^F$ | $p_t^F$ | $w_t^F$ | $b_t^F$ | $a_t^F$ | $r_t^F$ | $p_t^F$ | $w_t^F$ | $b_t^F$ |
| *(-,m-3s)* | 0 | 0 | 2 | 0 | 0 | 0 | 0 | 0 | 0 | 0 |
| *[m-3s,m-2s)* | 3 | 1 | 8 | 1 | 2 | 0 | 0 | 2 | 0 | 0 |
| *[m-2s,m-s)* | 10 | 28 | 17 | 15 | 13 | 2 | 10 | 5 | 3 | 7 |
| *[m-s,m)* | 67 | 63 | 44 | 60 | 76 | 28 | 19 | 20 | 28 | 25 |
| *[m,m+s)* | 64 | 58 | 66 | 71 | 71 | 18 | 17 | 19 | 17 | 13 |
| *[m+s,m+2s)* | 20 | 19 | 37 | 23 | 11 | 5 | 8 | 9 | 6 | 10 |
| *[m+2s,m+3s)* | 7 | 4 | 0 | 2 | 0 | 1 | 2 | 1 | 1 | 0 |
| *[m+3s,-)* | 1 | 1 | 0 | 2 | 1 | 1 | 0 | 0 | 1 | 1 |
| *[m-s,m+s]* | 76.16% | 69.54% | 63.22% | 75.29% | 84.48% | 83.64% | 64.29% | 69.64% | 80.36% | 67.86% |
| *[m-2s,m+2s]* | 93.60% | 96.55% | 94.25% | 97.13% | 98.28% | 96.36% | 96.43% | 94.64% | 96.43% | 98.21% |
| *[m-3s,m+3s]* | 99.42% | 99.43% | 98.85% | 98.85% | 99.43% | 98.18% | 100.00% | 100.00% | 98.21% | 98.21% |

Figure 3: Frequency histograms of the components (*m*=mean, *s*=standard deviation)

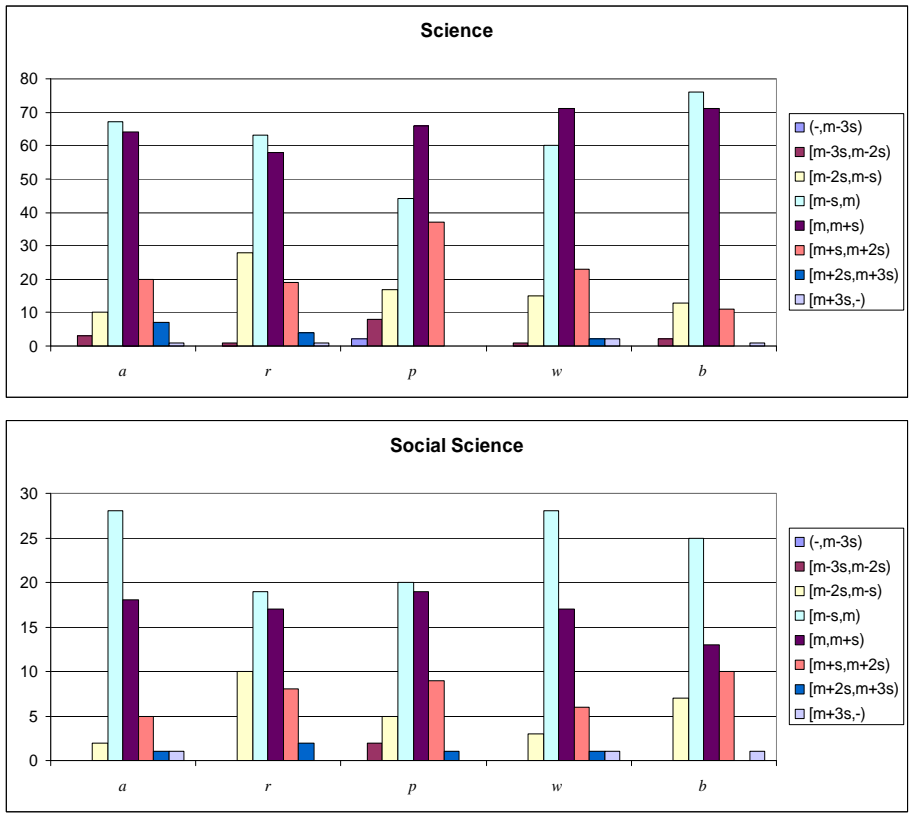

Figure 4: CNIF of all journals in the economics and business field

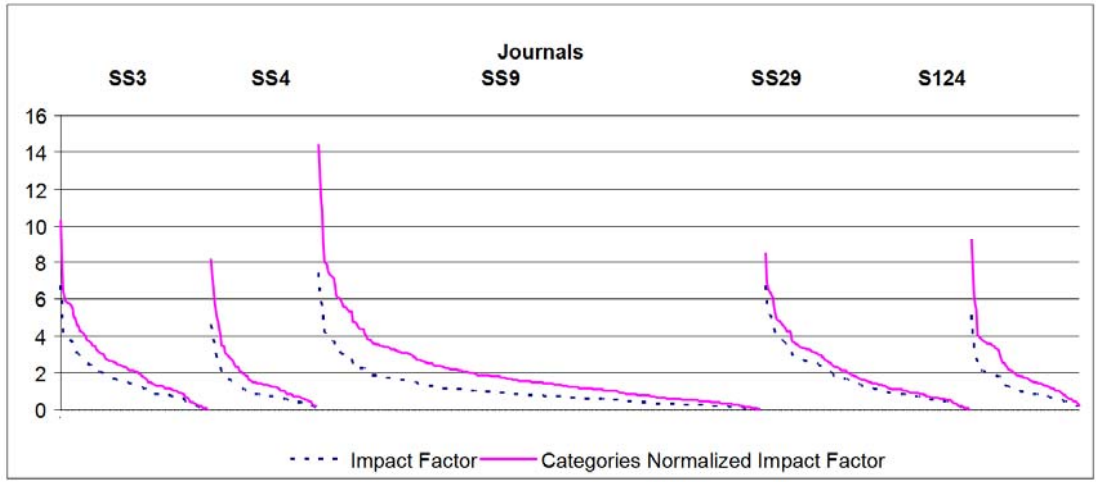

Figure 5: CNIF of all journals in two or more categories of the economics and business field

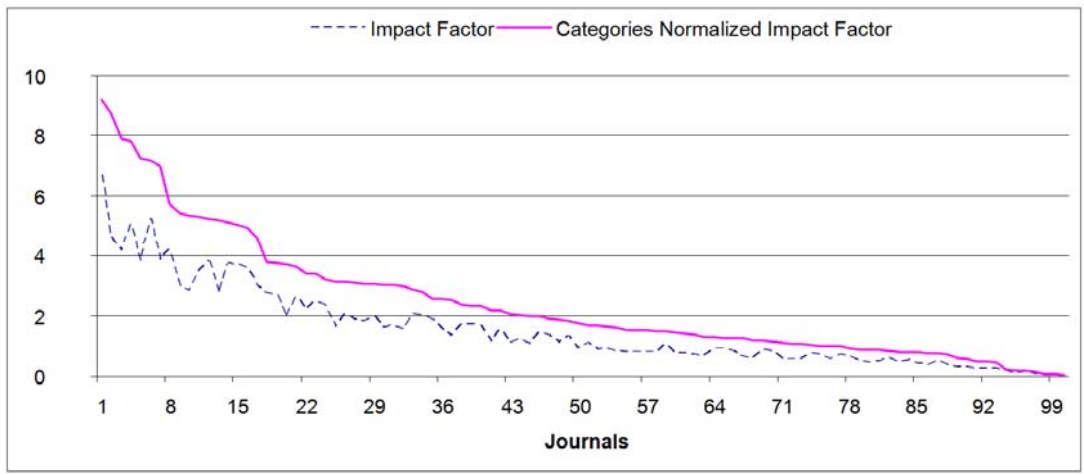